\documentclass[prd,twocolumn,showpacs,preprintnumbers,amsmath,amssymb,superscriptaddress,floatfix,nofootinbib]{revtex4}

\usepackage{graphicx}
\usepackage{amsmath}
\usepackage{amsfonts}
\usepackage{amssymb}
\usepackage{color}
\usepackage{multirow}
\usepackage[colorlinks, citecolor=blue,anchorcolor=red,menucolor=red,linkcolor=red,filecolor=red,runcolor=red,urlcolor=blue,frenchlinks=red]{hyperref}

\usepackage{subfigure}

\begin{document}

\title{The resonances $X(4140)$, $X(4160)$, and $P_{cs}(4459)$ in the decay of $\Lambda_b\to J/\psi\Lambda\phi$}

\author{Wen-Ying Liu}
\affiliation{School of Physics and Microelectronics, Zhengzhou University, Zhengzhou, Henan 450001, China}

\author{Wei Hao}
\affiliation{CAS Key Laboratory of Theoretical Physics, Institute of Theoretical Physics, Chinese Academy of Sciences, Beijing 100190,China}
\affiliation{University of Chinese Academy of Sciences (UCAS), Beijing 100049, China}

\author{Guan-Ying Wang}
\affiliation{School of Physics and Electronics, Henan University, Kaifeng 475004, China}

\author{Yan-Yan Wang}
\affiliation{School of Materials Science and Engineering, Zhengzhou University of Aeronautics, Henan 450001, China}

\author{En Wang}\email{wangen@zzu.edu.cn}
\affiliation{School of Physics and Microelectronics, Zhengzhou University, Zhengzhou, Henan 450001, China}

\author{De-Min Li}\email{lidm@zzu.edu.cn}
\affiliation{School of Physics and Microelectronics, Zhengzhou University, Zhengzhou, Henan 450001, China}

\begin{abstract}
We study the decay of $\Lambda_b\to J/\psi\Lambda\phi$ by taking into account the intermediate resonances $X(4140)$,  $X(4160)$, and  $P_{cs}(4459)$.  In addition to the peak of the $X(4140)$, we also find a bump structure around 4160~MeV followed by a cusp structure around $D^*_s\bar{D}^*_s$ threshold in the $J/\psi\phi$ invariant mass distribution of the  $\Lambda_b\to J/\psi\Lambda\phi$ process, which can be associated to the $D^*_s\bar{D}^*_s$ molecular state $X(4160)$. We also show that this reaction can also be used to confirm the existence of the hidden-charm pentaquark $P_{cs}(4459)$ with strangeness. We predict an enhancement structure close to the threshold in the $D^*_s\bar{D}^*_s$ invariant mass distribution of the  $\Lambda_b\to D^*_s\bar{D}^*_s\Lambda$ process, which is the reflection of the $X(4160)$ and should not be misidentified with a new resonance. We strongly suggest experimentalists to further measure these processes, which can be useful to clarify the nature of $X(4140)$ and $X(4160)$ resonances, and to confirm the existence of the $P_{cs}(4459)$.
\end{abstract}



\maketitle

\section{INTRODUCTION}
\label{sec:INTRODUCTION}
In the last decades, a lot of exotic multi-quark states were observed experimentally, which demonstrates that there exist more complex configurations than the conventional mesons and baryons predicted by the classical quark model~\cite{Brambilla:2019esw,Chen:2016qju,Liu:2019zoy}. Many of them are reported in the weak decays of the heavy hadrons, and the theoretical studies on the weak decays of the heavy hadrons have also increased our understanding of hadron-hadron interactions and the nature of many intermediate exotic states~\cite{Guo:2017jvc,Oset:2016lyh}.

It should be stressed that the weak decays of the $\Lambda_b$ have been proved to be a rich source of exotic spectroscopy, such as the new pentaquark states $P_c(4312)^+$, $P_c(4380)^+$, and $P_c(4350)^+$ observed by the LHCb Collaboration~\cite{Aaij:2015tga,Aaij:2019vzc}.
Very recently, the CMS Collaboration has reported the observation of the $\Lambda_b \to J/\psi\Lambda\phi$ decay using proton-proton collision data at the LHC, the ratio of the branching fraction $\mathcal{B}(\Lambda_b \to J/\psi\Lambda\phi)/{\mathcal{B}(\Lambda_b \to \psi(2s)\phi)} = (8.26 \pm 0.90 \pm 0.68 \pm 0.11) \times {10}^{-2} $~\cite{Sirunyan:2019dwp}. The observation of this reaction has opened a window on future searches for new resonances in the $J/\psi\Lambda$ and $J/\psi\phi$ systems, once a sufficient number of signal events is accumulated, as mentioned by the CMS Collaboration~\cite{Sirunyan:2019dwp}.

A series of meson-baryon interactions in the hidden charm sector, including the $J/\psi\Lambda$, were studied within the framework of the coupled channel unitary approach with the local hidden gauge formalism in Ref.~\cite{Wu:2010vk}, and two hidden charmonium pentaquark states were predicted in the isospin $I=0$ and strangeness $S=-1$ sector. Since the predicted hidden charm $S=-1$ pentaqurak couples in $S$-wave to $J/\psi\Lambda$ with a reasonable strength, some reactions involving the $J/\psi\Lambda$ final state system are proposed to search for the hidden charmonium pentaquark state, such as $\Lambda_b\to J/\psi K^0\Lambda$~\cite{Lu:2016roh}, $\Xi_b\to J/\psi K^-\Lambda$~\cite{Chen:2015sxa}, $\Lambda_b\to J/\psi \eta \Lambda$~\cite{Feijoo:2015kts}.
There are also some studies about the hidden-charm pentaquark state with strangeness~\cite{Anisovich:2015zqa,Wang:2015wsa,Chen:2016ryt,Xiao:2019gjd,Zhang:2020cdi,Shen:2020gpw}.
Basing the suggestion proposed in Ref.~\cite{Chen:2015sxa}, the LHCb has started to search for the hidden-charm pentaquark with strangeness in Ref.~\cite{Aaij:2017bef}.
Very recently, the LHCb Collaboration reported  a hidden-charm pentaquark state with strangeness $P_{cs}(4459)$ in the $J/\psi\Lambda$ invariant mass spectrum of the $\Xi^-_b\to J/\psi K^-\Lambda$ process, and the mass and width of $P_{cs}(4459)$ were measured to be~\cite{LHCbPcs}
\begin{eqnarray}
M_{P_{cs}}&=& 4458.8\pm2.9 ^{+4.7}_{-1.1}~{\rm MeV},\nonumber \\
\Gamma_{P_{cs}}&=& 17.3\pm 6.5 ^{+8.0}_{-5.7}~{\rm MeV}.
\end{eqnarray}
Although the quantum numbers of the $P_{cs}(4459)$ have not been determined, a hidden-charm pentaquark with strangeness $S=-1$ and $J^P=1/2^-$ was predicted to have a mass much close to the $P_{cs}(4459)$ mass in many works~\cite{Chen:2016ryt,Xiao:2019gjd,Chen:2020uif}. Thus, we will take into account the contribution from the $P_{cs}(4459)$ in the $J/\psi\Lambda$ system of the $\Lambda_b\to J/\psi\Lambda \phi$ process.

For the $J/\psi \phi$ system, one of the most important resonances is the charmonium-like state $X(4140)$, which was first reported in the $B^+\to J/\psi\phi K^+$ process by the CDF Collaboration~\cite{Aaltonen:2009tz}, and later confirmed by the CMS~\cite{Chatrchyan:2013dma}, D0~\cite{Abazov:2013xda,Abazov:2015sxa}, and LHCb Collaborations~\cite{Aaij:2016iza,Aaij:2016nsc}. However, the width of the $X(4140)$ is measured to be $83\pm21^{+4.6}_{-2.8}$~MeV by LHCb~\cite{Aaij:2016iza,Aaij:2016nsc}, much larger than the data from the other experiments~\cite{Zyla:2020zbs}, which confuses the understanding of the $X(4140)$ nature~\cite{Hao:2019fjg,Chen:2016iua}. We have provided a reasonable explanation of the LHCb measurements~~\cite{Aaij:2016iza,Aaij:2016nsc} by considering the contributions from the $X(4160)$ and the  $X(4140)$ with a narrow width, and have predicted a broad bump around 4160~MeV followed by a cusp at $D^*_s\bar{D}^*_s$ threshold in the $J/\psi\phi$ invariant mass distribution, which is one feature of the $D^*_s\bar{D}^*_s$ molecular state $X(4160)$~\cite{Wang:2017mrt}.
In addition, the predicted $J/\psi\phi$ invariant mass distribution of the $e^+e^-\to \gamma J/\psi \phi$ involving the intermediate $X(4140)$ and $X(4160)$ are also compatible with the BESIII measurement~\cite{Wang:2018djr}.

The $S$-wave interaction of ten vector-vector channels with charm $C=0$ and strangeness $S=0$ ($J/\psi\phi$, $D_{s}^{*}\bar{D}_{s}^{*}$, $D^*\bar{D}^*$, $K^*\bar{K}^*$, $\rho\rho$, $\omega\omega$, $\phi\phi$, $J/\psi J/\psi$, $\omega J/\psi$, $\omega\phi$) were studied within the hidden gauge formalism, and one $D^*\bar{D}^*$ molecular state with $I(J^{PC})=0(2^{++})$ was predicted, which could be identified as $X(4160)$~\cite{Molina:2009ct}.
Based on the above discussions, we expect that the $X(4140)$ and $X(4160)$ play an important role in the reaction of $\Lambda_b\to J/\psi \Lambda \phi$.

In this work, we will study the reaction of $\Lambda_b\to J/\psi \Lambda \phi$ by taking into account the contributions from the $X(4140)$,  $X(4160)$, and  $P_{cs}(4459)$, and also propose to test the molecular nature of the $X(4160)$ by looking for the enhancement structure near the $D^*_s\bar{D}^*_s$ threshold in the $\Lambda_b\to D{_{s}^{*}}\bar{D}{_{s}^{*}}\Lambda$ reaction, since more data samples of those two processes are crucial to learn more about the charmonium-like states $X(4140)$ and $X(4160)$ and especially to confirm the existence of the recently observed $P_{cs}(4459)$. 

This paper is organized as follows. In Sec.~\ref{sec:FORMALIISM}, we will  present the mechanism for the reaction of $\Lambda_b\to J/\psi\Lambda\phi$ and $\Lambda_b\to D{_{s}^{*}}\bar{D}{_{s}^{*}}\Lambda$, and in Sec.~\ref{sec:RESULTS}, we will show our results and discussions, followed by a short summary in the last section.

\section{FORMALISM}
\label{sec:FORMALIISM}

\subsection{The $\Lambda_b\to J/\psi\Lambda\phi$ process}

\begin{figure}[htbp]
\centering
\subfigure[]{\includegraphics[width=6.5cm]{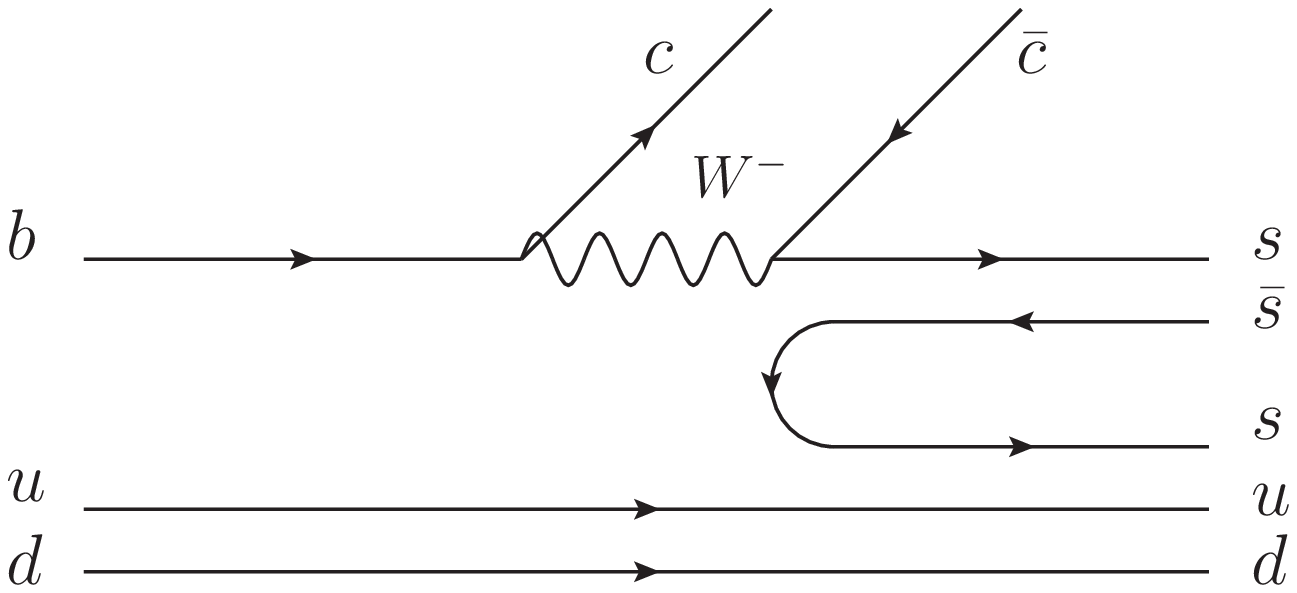}\label{fig:1a}}
\subfigure[]{\includegraphics[width=6.5cm]{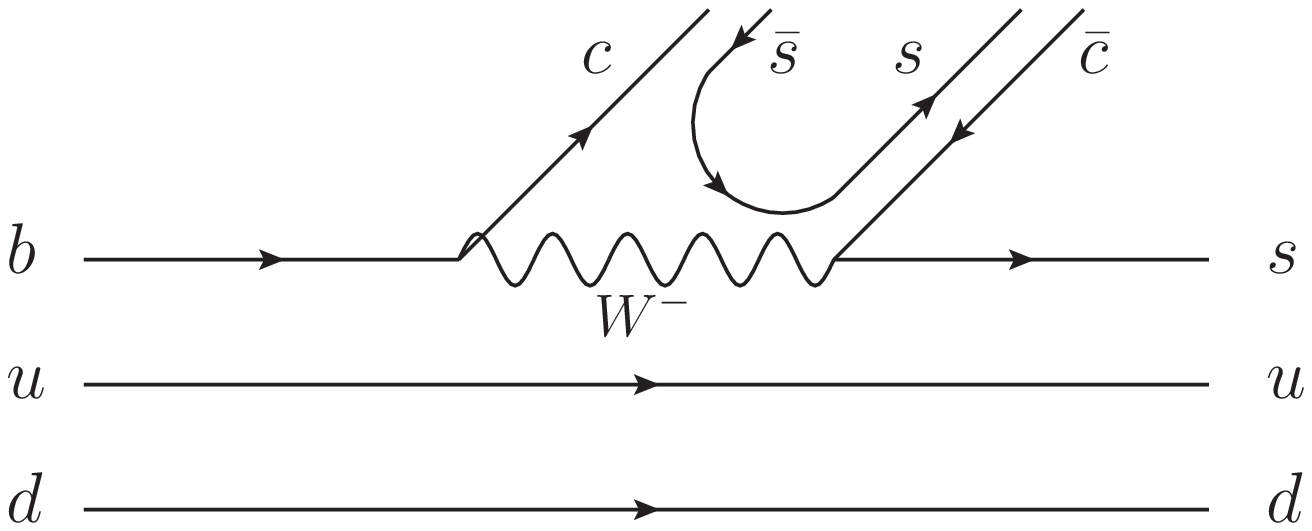}\label{fig:1b}}
\subfigure[]{\includegraphics[width=6.5cm]{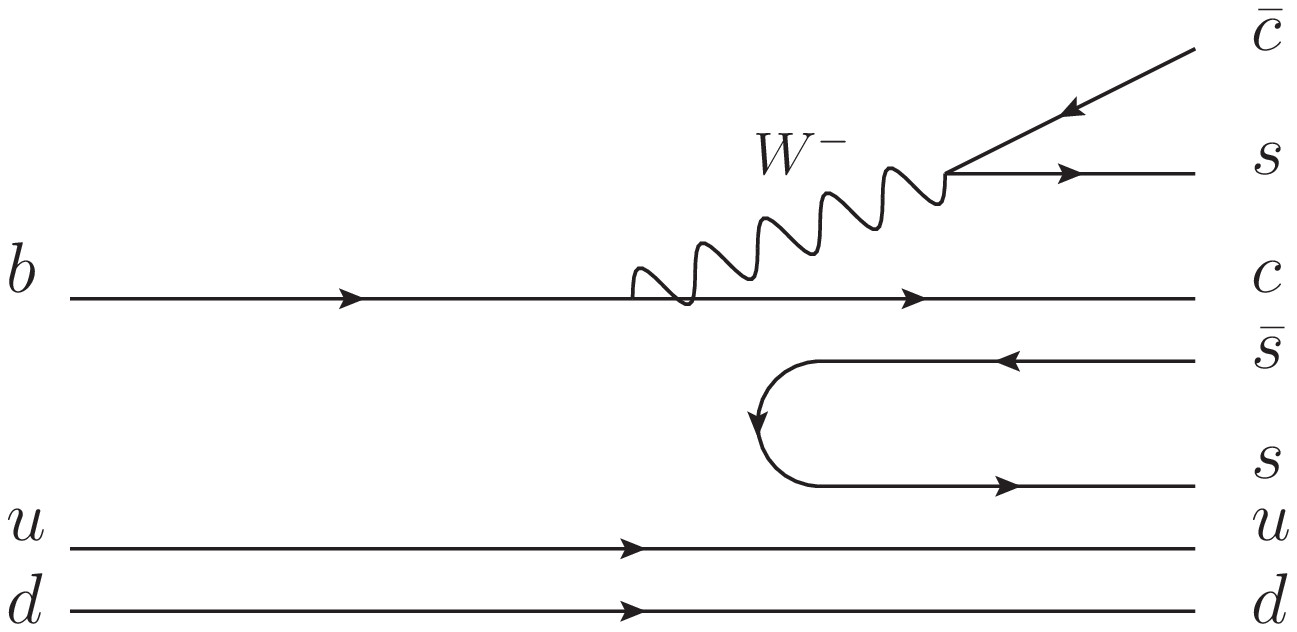}\label{fig:1c}}
\caption{The $\Lambda_b$ weak decay at the microscopic quark picture. (a) The $\Lambda{_b}\to J/\psi\Lambda\phi$ decay through the internal $W^-$ emission and hadronization of $sud$ through $q\bar{q}$ created with vacuum quantum numbers; (b) the $\Lambda_b\to D{_{s}^{*}}\bar{D}{_{s}^{*}}\Lambda$ decay through the internal $W^-$ emission and hadronization of $c\bar{c}$ through $\bar{s}s$; (c)  the $\Lambda_b\to D{_{s}^{*}}\bar{D}{_{s}^{*}}\Lambda$ decay through the external $W^-$ emission and hadronization of $cud$  with $\bar{s}s$.}
\label{fig:1}
\end{figure}

Analogous to the Refs.~\cite{Lu:2016roh,Chen:2015sxa,Feijoo:2015kts,Wang:2017mrt,Sirunyan:2019dwp,Magas:2020zuo},  the $\Lambda_b\to J/\psi\Lambda\phi$ process can perform through the $W^-$ internal emission mechanism. In the first step, the $b$ quark of the initial $\Lambda_b$ weakly decays into a $c$ quark and a $W^-$ boson, followed by the $W^- $ decaying into $\bar{c}s$  pair, which can be expressed as,
\begin{eqnarray}
\left|\Lambda_b\right\rangle &=& \frac{1}{\sqrt{2}} b (ud-du)  \nonumber \\
&\Rightarrow & V_{bc} \frac{1}{\sqrt{2}} c W^- (ud-du)  \nonumber \\
&\Rightarrow & V_{bc} V_{cs} \frac{1}{\sqrt{2}} c \bar{c}s (ud-du) , \label{eq:weakdecay}
\end{eqnarray}
where we take the flavor wave function $\Lambda_b\ =b(ud-du)/\sqrt{2}$,  $V_{bc}$ and $V_{cs}$ are the elements of the Cabibbo-Kobayashi-Maskawa (CKM) matrix. In order to give rise to the final states of this process, the quark components of Eq.~(\ref{eq:weakdecay}), together with an extra $\bar{q}q$ ($\equiv \bar{u}u + \bar{d}d +\bar{s}s$) created from the vacuum, should hadronize into hadrons, as depicted in Fig.~\ref{fig:1a},
\begin{eqnarray}
\left| H^{(a)} \right\rangle &=& V_{bc}V_{cs} J/\psi\left| s ( \bar{s}s) \frac{(ud-du)}{\sqrt{2}} \right\rangle \nonumber \\
&=&  V_{bc}V_{cs}  J/\psi \phi \Lambda.
\end{eqnarray}

On the other hand, since the $X(4160)$ was predicted to strongly couple to the $D^*_s\bar{D}^*_s$ channel and weakly couple to the $J/\psi\phi$ channel~\cite{Molina:2009ct},  the $c\bar{c}sud$ component from the $\Lambda_b$ weak decay, together with the extra $\bar{s}s$ created from the vacuum, could hadronize to the  $D^*_s\bar{D}^*_s\Lambda$ in two ways: the $W^-$ internal emission of Fig.~\ref{fig:1b} and the $W^-$ external emission of Fig.~\ref{fig:1c}. Although this is not the final state of $J/\psi\Lambda\phi$ that one observes, the idea is that the $D{_{s}^{*}}\bar{D}{_{s}^{*}}$ will interact and create a resonance that couples to $J/\psi\phi$. We have,
\begin{eqnarray}
\left| H^{(b)} \right\rangle &=& V_{bc}V_{cs} \left| c (\bar{s}s) \bar{c} s \frac{(ud-du)}{\sqrt{2}} \right\rangle \nonumber \\
&=&  V_{bc}V_{cs}  D^*_s\bar{D}^*_s \Lambda,\\
\left| H^{(c)} \right\rangle &=& V_{bc}V_{cs} \left|  \bar{c} s c (\bar{s} s ) \frac{(ud-du)}{\sqrt{2}} \right\rangle \nonumber \\
&=&  V_{bc}V_{cs} \bar{D}^*_s D^*_s \Lambda,
\end{eqnarray}
which finally produce $J/\psi\phi$ via the $D_s^*\bar{D}_s^*$ final state interaction.

Although we can straightly get the final state of $J/\psi\phi\Lambda$ via the mechanism of Fig.~\ref{fig:1a}, this mechanism is penalized by a color factor with respect to the $W^-$ external emission of Fig.~\ref{fig:1c}, and this term with intermediate state $J/\psi\phi$ instead of $D{_{s}^{*}}\bar{D}{_{s}^{*}}$ would involve the extra factor ${g_{J/\psi\phi}}/{g_{D{_{s}^{*}}\bar{D}{_{s}^{*}}}}=|-2617-i5151|/|18927-i5524|\approx 0.3$ versus the amplitudes of Figs.~\ref{fig:1b} and \ref{fig:1c}, where ${g_{J/\psi\phi}}=(-2617-i515)$~MeV and ${g_{D{_{s}^{*}}\bar{D}{_{s}^{*}}}}=(18927-i5524)$~MeV are the couplings of the $D^*_s\bar{D}^*_s$ molecule to $J/\psi\phi$ and $D^*_s\bar{D}^*_s$~\cite{Molina:2009ct}. Thus, we will neglect the contribution from the Fig.~\ref{fig:1a} in this work.

\begin{figure}[htbp]
\centering
\subfigure[]{\includegraphics[width=6.5cm]{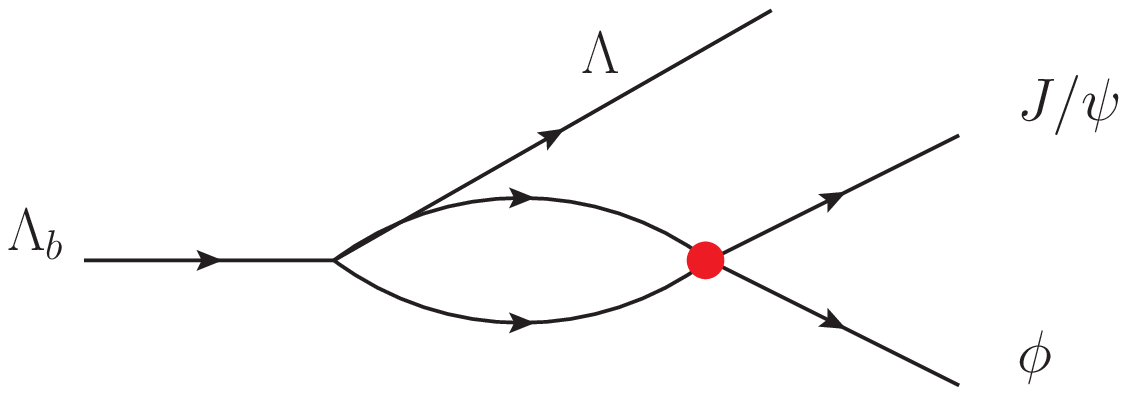}\label{fig:2a}}
\subfigure[]{\includegraphics[width=6.5cm]{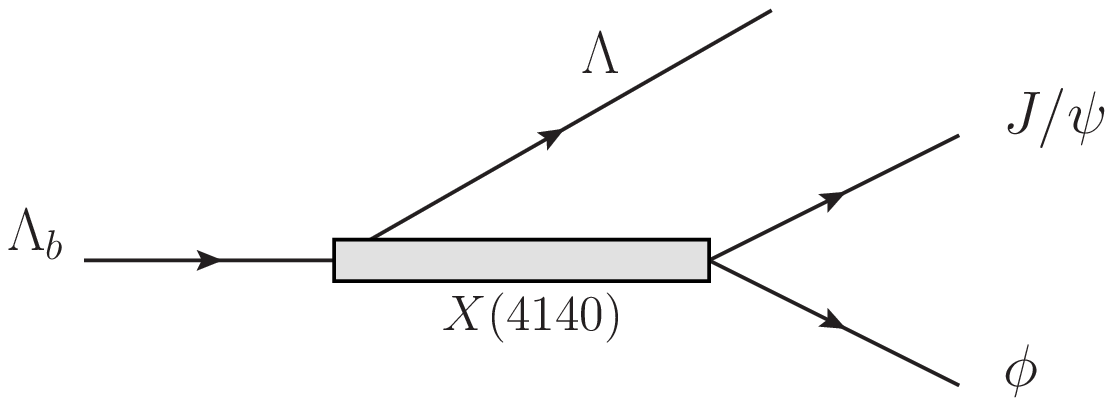}\label{fig:2b}}
\subfigure[]{\includegraphics[width=6.5cm]{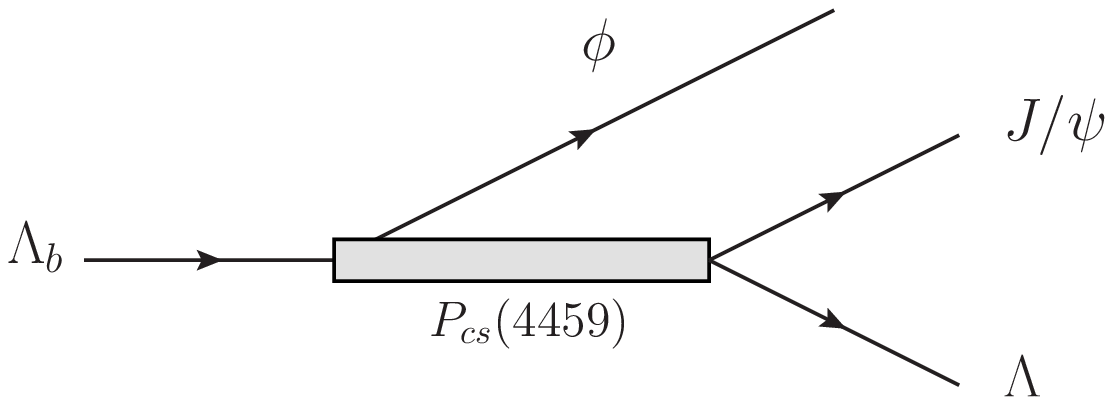}\label{fig:2c}}
\caption{\small{Mechanism to produce the $J/\psi\Lambda\phi$ final state through $X(4160)$, $X(4140)$, and $P_{cs}(4459)$}.}
\label{fig:2}
\end{figure}

Now, we need the final state interaction of $D_s^*\bar{D}_s^*$ produced in Figs.~\ref{fig:1b} and \ref{fig:1c} to give rise to the $J/\psi\phi$ final state, and also generate dynamically the $D_s^*\bar{D}_s^*$ molecular state $X(4160)$, as shown in Fig.~\ref{fig:2a}.
Since the quantum numbers of $X(4160)$ are predicted to be $J^{PC}=2^{++}$ in Refs.~\cite{Molina:2009ct,Torres:2016oyz}, the $\Lambda$ should be in $P$-wave to match to the spin $J=1/2$ of the initial $\Lambda_b$, and the amplitude can be expressed as,
\begin{equation}
\mathcal{M}^P = A\left(\vec{\epsilon}_{J/\psi}\times\vec{\epsilon}_\phi\right)\cdot \vec{k}\,G_{D{_s^*}\bar{D}{_s^*}}\, t_{D{_s^*}\bar{D}{_s^*}\to J/\psi\phi},
\label{eq1}
\end{equation}
where the unknown constant $A$ represents the total weight of the microscopic decay in the quark level showed in Figs.~\ref{fig:1b} and \ref{fig:1c}; $\vec{\epsilon}_{J/\psi}$ and $\vec{\epsilon}_\phi$ are the polarization vectors  of the $J/\psi$ and $\phi$, respectively; $\vec{k}$ is the $\Lambda$ momentum in the rest frame of the $J/\psi\phi$ system;
$G_{D{_s^*}\bar{D}{_s^*}}$ is the loop function of the $D{_s^*}\bar{D}{_s^*}$ channel, and $t_{D{_s^*}\bar{D}{_s^*}\to J/\psi\phi}$ is the transition amplitude of $D{_s^*}\bar{D}{_s^*}\to J/\psi\phi$, both of which are the functions of the $J/\psi\phi$ invariant mass $M_{J/\psi\phi}$.
To avoid potential dangers using the dimensional regularizations as pointed out in Ref.~\cite{Wu:2010rv},
in this work we use the cut-off method and  take $q_{\rm max}=630$~MeV followed Ref.~\cite{Wang:2018djr},
\begin{eqnarray}
G_{D{_s^*}\bar{D}{_s^*}}&=& i \int{\frac{d^{4}q}{(2\pi)^4}\frac{1}{q^2-m_1^2+i\epsilon}\frac{1}{(p-q)^2-m_2^2+i\epsilon}} \nonumber \\
 &=& \int_{0}^{q_{\rm max}}{\frac{q^2dq}{(2\pi)^2}\frac{\omega_1+\omega_2}{\omega_1\omega_2[(P^0)^2-(\omega_1+\omega_2)^2+i\epsilon]}}, \nonumber  \\
\label{eq:loop}
\end{eqnarray}
where $m_{1,2}$ are the masses of $D^*_s$ and $\bar{D}^*_s$, $\omega_i=\sqrt{|\vec{q}_i|^2+m^2_i}$ are the energies of $D^*_s$ and $\bar{D}^*_s$, and $P^0=\sqrt{s}$ is the center-of-mass energy.
And the transition amplitude is given by,
\begin{equation}
t_{D{_s^*}\bar{D}{_s^*}\to J/\psi\phi}= \frac{g_{J/\psi\phi}g_{D{_s^*}\bar{D}{_s^*}}}{M_{J/\psi\phi}^{2}-M_{X(4160)}^2 + i M_{X(4160)} \Gamma_{X(4160)}}
\label{eq2}
\end{equation}
with the couplings $g_{D^*_s\bar{D}^*_s}$ and  $g_{J/\psi\phi}$ obtained in Ref.~\cite{Molina:2009ct}, the $X(4160)$ mass $M_{X(4160)}=4156$~MeV~\cite{Zyla:2020zbs}.  For the $X(4160)$ width $\Gamma_{X(4160)}$, we take into account the
Flatt{\'e} effect for the opening of the important $D^*_s\bar{D}^*_s$ channel above the $D^*_s\bar{D}^*_s$ threshold as follows,
\begin{eqnarray}
\Gamma_{X(4160)}&=&\Gamma_0+\Gamma_{J/\psi\phi}+\Gamma_{D{_s^*}\bar{D}{_s^*}} ,\\
\Gamma_{J/\psi\phi}&=& \frac{{|g_{J/\psi\phi}|}^2\,\widetilde{p}_{\phi}}{8\pi {M_{X(4160)}^2}},  \\
\Gamma_{D{_s^*}\bar{D}{_s^*}}&=& \frac{{|g_{D{_s^*}\bar{D}{_s^*}}|}^2\,  \widetilde{p}_{D_s^*}}{8\pi {M_{X(4160)}^2}} \Theta \left(M_{J/\psi\phi}-2m_{D^*_s}\right),\\
\widetilde{p}_{\phi}&=&\frac{\lambda^{1/2}(M_{J/\psi\phi}^2,m_{J/\psi}^2,m_{\phi}^2)}{2M_{J/\psi\phi}},\\
\widetilde{p}_{D_s^*}&=&\frac{\lambda^{1/2}(M_{J/\psi\phi}^2,m_{D_s^*}^2,m_{\bar{D}_s^*}^2)}{2M_{J/\psi\phi}},
\end{eqnarray}
where $\Gamma_0$ is taken as $65.0$~MeV by fitting to the LHCb data of the $B^+\to J/\psi\phi K^+$ process~\cite{Wang:2017mrt}, $\widetilde{p}_{\phi}$ and $\widetilde{p}_{D_s^*}$ are respectively the momenta of $\phi$ and $D_s^*$ in the rest frame of $J/\psi\phi$. $\lambda(x,y,z)=x^2+y^2+z^2-2xy-2xz-2yz$ is the K{\"a}llen function.

On the other hand, the $X(4140)$ state was observed in the $J/\psi\phi$ invariant mass distribution of the process $B^+\to J/\psi\phi K^+$~\cite{Zyla:2020zbs}, and could be interpreted as the $\chi_{c1}(3P)$ state~\cite{Hao:2019fjg,Chen:2016iua}, thus the decay $\Lambda_b\to J/\psi \phi \Lambda$ can also happen through the intermediate resonance $X(4140)$ with $\Lambda$ in $S$-wave, and the corresponding amplitude can be expressed as,
\begin{equation}
\mathcal{M}^S=B \times \frac{M_{X(4140)}^3\, \vec{\epsilon}_{J/\psi}\cdot \vec{\epsilon}_\phi}{M_{J/\psi\phi}^{2}-M_{X(4140)}^2 + i M_{X(4140)}\Gamma_{X(4140)}},
\label{eq5}
\end{equation}
where the parameter $B$, with the same dimension as $A$, accounts for the relative weight of the contribution from the intermediate resonance $X(4140)$. We take $M_{X(4140)}=4135$~MeV and $\Gamma_{X(4140)}=19$~MeV~~\cite{Zyla:2020zbs}. Since the $\Lambda_b\to J/\psi \phi \Lambda$ process has been observed by the CMS Collaboration~\cite{Sirunyan:2019dwp}, we will show our results up to an arbitrary normalization, and our predictions can be tested while enough data sample are available in future.

As we discussed above, many studies favor the quantum numbers $J^P=1/2^-$ for the $P_{cs}(4459)$~\cite{Chen:2020uif,Peng:2020hql,Wang:2020eep}, which implies that the $P_{cs}(4459)$ couples to $J/\psi\Lambda$ in $S$-wave, and the $\Lambda_b$ weakly decays into $P_{cs}(4459)\phi$ in $S$-wave, as depicted in Fig.~\ref{fig:2c}. The amplitude for the contribution of $P_{cs}(4459)$ can be expressed as,
\begin{eqnarray}
\mathcal{M}^{P_{cs}}=C \times \frac{M_{P_{cs}}^3\, \vec{\epsilon}_{J/\psi}\cdot \vec{\epsilon}_\phi}{M_{J/\psi\Lambda}^{2}-M_{P_{cs}}^2 + i M_{P_{cs}}\Gamma_{P_{cs}}},
\label{eq:ampPcs}
\end{eqnarray}
where we take $M_{P_{cs}}=4458.8$~MeV and $\Gamma_{P_{cs}}=17.3$~MeV~\cite{LHCbPcs}, and the parameter $C$ stands for the relative weight of the contribution from the $P_{cs}(4459)$.

With the amplitudes of Eqs.~(\ref{eq1}), (\ref{eq5}) and (\ref{eq:ampPcs}), the mass distribution of the $\Lambda_b\to J/\psi\phi\Lambda$ process is given by,
\begin{eqnarray}
\frac{d^2\Gamma}{dM^2_{J/\psi\phi}dM^2_{J/\psi\Lambda}}&=& \frac{1}{(2\pi)^3}\frac{1}{32M^3_{\Lambda_b}} \sum |\mathcal{M}|^2,
\end{eqnarray}
with
\begin{eqnarray}
\sum |\mathcal{M}|^2 &=& \sum \left( |\mathcal{M}^S|^2+|\mathcal{M}^P|^2\right) \label{eq:totalamp} \\
&=& B^2 \left( 3 |\tilde{\mathcal{M}}^S|^2+2|\vec{k}|^2|\tilde{\mathcal{M}}^P|^2\right), \label{eq:dw} \\
\tilde{\mathcal{M}}^P &=&\alpha G_{D{_s^*}\bar{D}{_s^*}}\, t_{D{_s^*}\bar{D}{_s^*}\to J/\psi\phi}, \\
\tilde{\mathcal{M}}^S&=& \frac{M_{X(4140)}^3}{M_{J/\psi\phi}^{2}-M_{X(4140)}^2 + i M_{X(4140)}\Gamma_{X(4140)}}\nonumber \\
&+& \frac{\beta M_{P_{cs}}^3}{M_{J/\psi\Lambda}^{2}-M_{P_{cs}}^2 + i M_{P_{cs}}\Gamma_{P_{cs}}},
\end{eqnarray}
where $\alpha=A/B$ and $\beta=C/B$ stand for the relative weights of the $X(4160)$ and $P_{cs}(4459)$ terms with respect to the $X(4140)$ term, and we take $\alpha=\alpha_0=2100$ in order to give roughly equal strengths for $X(4140)$ and $X(4160)$, as done in Refs.~\cite{Wang:2017mrt,Magas:2020zuo}. It should be pointed out that there is no interference between the $S$-wave and $P$-wave, and the sum over the $J/\psi$ and $\phi$ polarizations is,
\begin{gather}
\sum_{\rm pol}\left[\left(\vec{\epsilon}_{J/\psi}\times\vec{\epsilon}_\phi\right)\cdot \vec{k} \right]^2 = 2\delta^{il} k^i  k^l =2 |\vec{k}|^2,  \\
\sum_{\rm pol}\left(\vec{\epsilon}_{J/\psi}\cdot\vec{\epsilon}_\phi\right)^2 =3.
\end{gather}

\subsection{The $\Lambda_b\to D{_{s}^{*}}\bar{D}{_{s}^{*}}\Lambda$ process}
\begin{figure}[htbp]
\centering
\subfigure[]{\includegraphics[width=6.5cm]{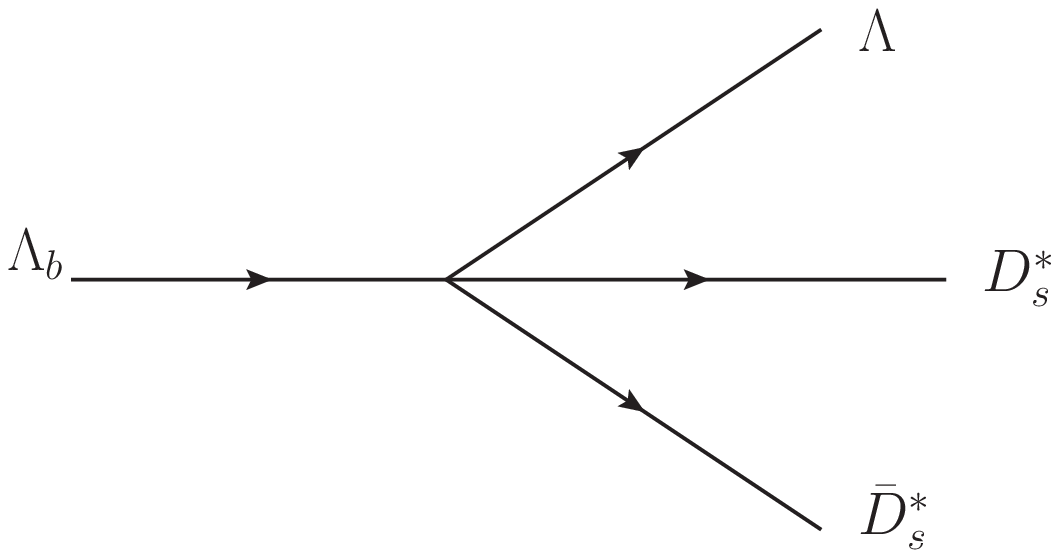}\label{fig:3a}}
\subfigure[]{\includegraphics[width=6.5cm]{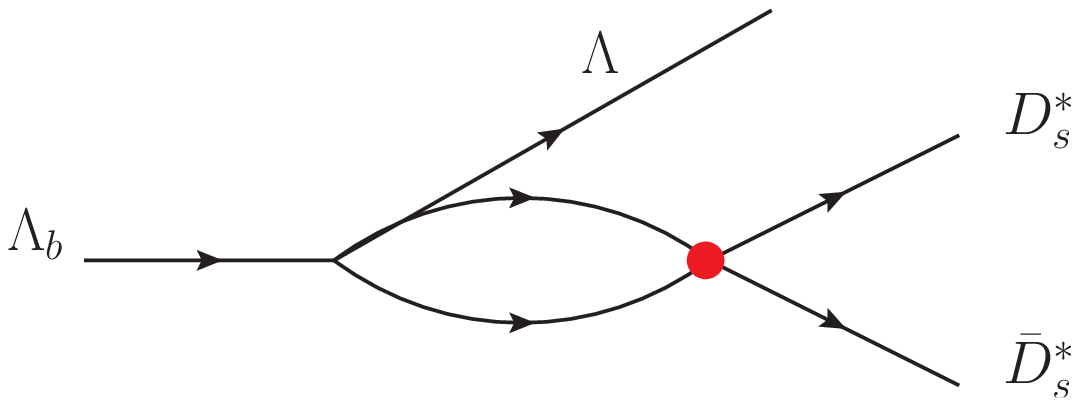}\label{fig:3b}}
\caption{\small{Mechanism to produce the $D{_s^*}\bar{D}{_s^*}$ final state}}
\label{fig:3}
\end{figure}

Since the width of the $X(4160)$ is $139^{+110}_{-60}$~MeV~\cite{Zyla:2020zbs}, and its mass is close to the one of the $X(4140)$, the signal of the bump structure from the $X(4160)$ is usually difficult to be identified. Because the $X(4160)$ is predicted to couple strongly to the $D^*_s\bar{D}^*_s$ channel, it is expected to give rise to an enhancement structure near  threshold in the $D^*_s\bar{D}^*_s$ invariant mass distribution. Thus, it is also interesting to investigate the $\Lambda_b\to D^*_s\bar{D}^*_s \Lambda$ process.  The mechanism of this process can also be depicted in Figs.~\ref{fig:1b} and \ref{fig:1c}, and we will take into account the contributions from the tree level diagram of Fig.~\ref{fig:3a} and the $D^*_s\bar{D}^*_s$ final state interaction of Fig.~\ref{fig:3b}. The amplitudes for both diagrams can be expressed as,
\begin{equation}
\mathcal{M}^S= D\times k_{\rm ave}\,\vec{\epsilon}_{D^*_s}\cdot\vec{\epsilon}_{\bar{D}^*_s}
\label{eq:swave}
\end{equation}
for $\Lambda$ in $S$-wave, and,
\begin{equation}
\mathcal{M}^P= E \left(\vec{\epsilon}_{D^*_s}\times\vec{\epsilon}_{\bar{D}^*_s}\right)\cdot \vec{k}^{\,\prime}\,\left(1+G_{D{_s^*}\bar{D}{_s^*}}\, t_{D{_s^*}\bar{D}{_s^*}\to D{_s^*}\bar{D}{_s^*}}\right)
\label{eq:pwave}
\end{equation}
for $\Lambda$ in $P$-wave, where $\vec{\epsilon}_{D^*_s}$ and $\vec{\epsilon}_{\bar{D}^*_s}$ are the polarization vectors of the $D^*_s$ and $\bar{D}^*_s$, respectively, and $\vec{k}^{\,\prime}$ is the $\Lambda$ momentum in the $D^*_s\bar{D}^*_s$ system.  We introduce $k_{\rm ave}=1000$~MeV to have the parameters $D$ and $E$ with same dimension. The loop function is given by Eq.~(\ref{eq:loop}), and the transition amplitude can be expressed as,
\begin{equation}
t_{D_s^*\bar{D}_s^*\to D_s^*\bar{D}_s^*}= \frac{g_{D_s^*\bar{D}_s^*}g_{D{_s^*}\bar{D}{_s^*}}}{M_{D_s^*\bar{D}_s^*}^{2}-M_{X(4160)}^2 + i M_{X(4160)} \Gamma_{X(4160)}}.
\label{eq:ampDsDs}
\end{equation}

Finally, the $D^*_s\bar{D}^*_s$ invariant  mass distribution of the $\Lambda_b\to D{_{s}^{*}}\bar{D}{_{s}^{*}}\Lambda$ process is given by,
\begin{eqnarray}
\frac{d\Gamma}{dM_{D_s^*\bar{D}_s^*}}&=& \frac{1}{(2\pi)^3}\frac{1}{4M^2_{\Lambda_b}} p_{\Lambda}\tilde{p}_{D^*_s} \sum |\mathcal{M}|^2 ,\label{eq:dw_dsds} \\
|\mathcal{M}|^2&=&D^2 \left(3 |k^2_{\rm ave}|+2|\vec{k}^{\,\prime}|^2  |\tilde{ \mathcal{M}}^P|^2 \right), \nonumber \\
\tilde{\mathcal{M}}^P&=&\gamma \left(1+G_{D{_s^*}\bar{D}{_s^*}}\, t_{D{_s^*}\bar{D}{_s^*}\to D{_s^*}\bar{D}{_s^*}}\right),
\label{eq9}
\end{eqnarray}
with $\gamma=E/D$. $p_\Lambda$ is the $\Lambda$ momentum in the rest frame of the $\Lambda_b$, $\tilde{p}_{D^*_s}$ is the $D^*_s$ momentum in the rest frame of the $D^*_s\bar{D}^*_s$ system.

\section{RESULTS AND DISCUSSIONS}
\label{sec:RESULTS}

\begin{figure}[htbp]
\centering
\includegraphics[width=7cm]{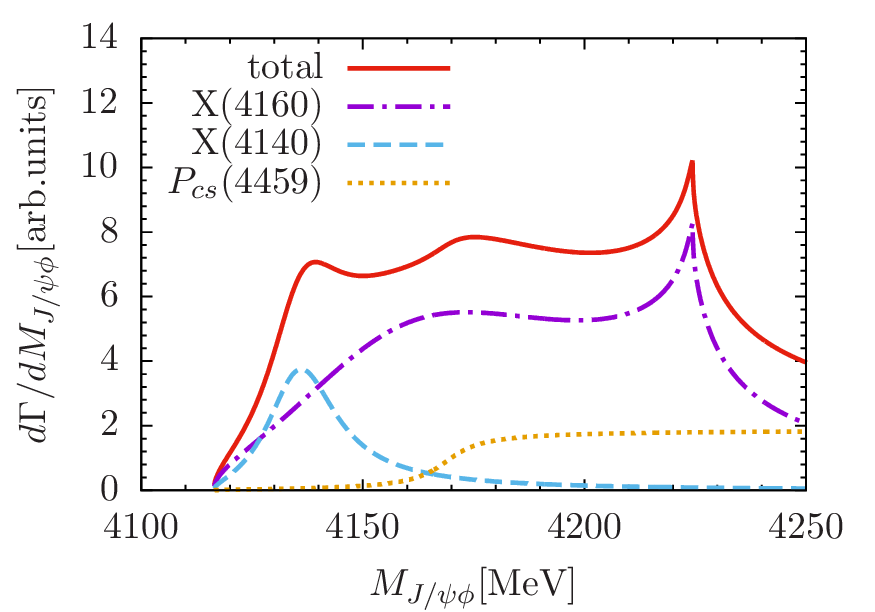}
\caption{The $J/\psi\phi$ invariant mass distribution of the $\Lambda_b\to J/\psi\Lambda\phi$ process. The curves labeled as `$X(4160)$', `$X(4140)$' and `$P_{cs}(4459)$' show the contributions from the the molecular state $X(4160)$, the intermediate resonances $X(4140)$ and $P_{cs}(4459)$, respectively. The curve labeled as `total' corresponds to the total contributions of Eq.~(\ref{eq:totalamp}).}
\label{fig:dw_jpsiphi}
\end{figure}

\begin{figure}[htbp]
\centering
\includegraphics[width=7cm]{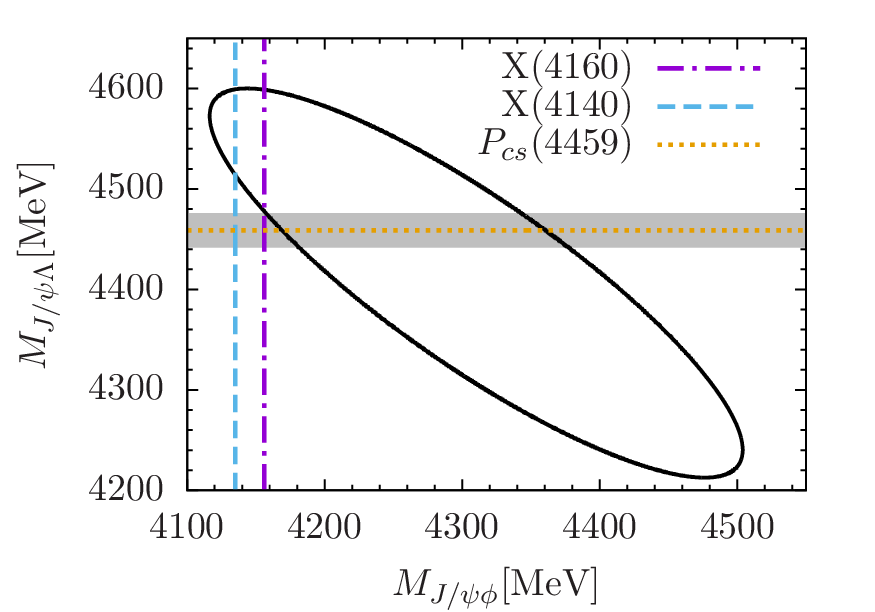}
\caption{The Dalitz plot of the $\Lambda_b\to J/\psi \Lambda \phi$ process.}
\label{fig:dalitz}
\end{figure}

\begin{figure}[htbp]
\centering
\includegraphics[width=6.5cm]{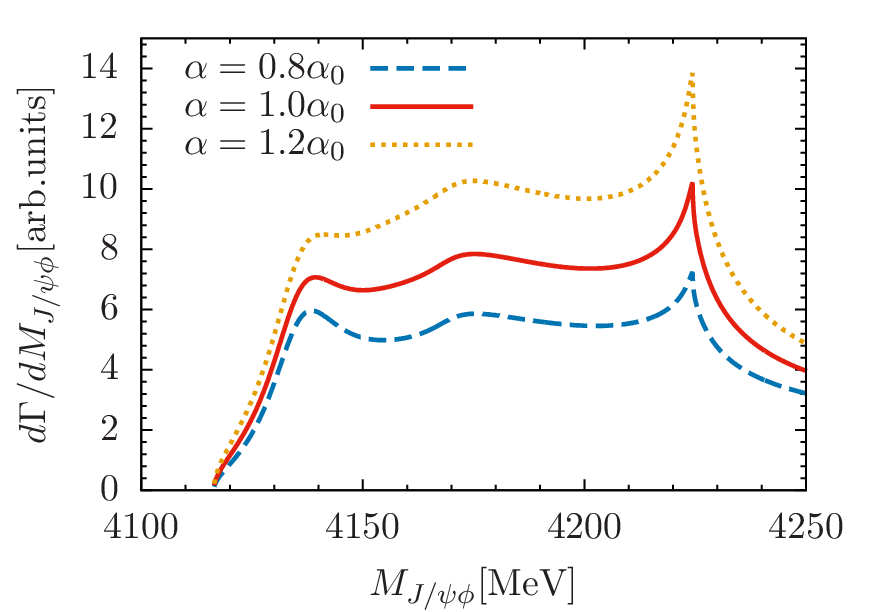}
\caption{The $J/\psi\phi$ invariant mass distribution of the $\Lambda_b\to J/\psi\phi\Lambda$ process with different values of $\alpha$ ($=0.8\alpha_0,~1.0\alpha_0,~1.2\alpha_0$).}
\label{fig:dw_jpsiphi_beta}
\end{figure}

In this section, we will present our results. With the $\beta=1$, we show the $J/\psi\phi$ mass distribution in Fig.~\ref{fig:dw_jpsiphi}. We can find  a peak around 4140~MeV, corresponding to $X(4140)$ resonance, and a bump around 4160~MeV followed by a cusp around $D{_{s}^{*}}\bar{D}{_{s}^{*}}$ threshold, which can be associated to the resonance $X(4160)$. It should be stressed that a molecular state that couples to several hadron-hadron channels may develop a strong and unexpected cusp in the invariant mass distribution in one of the weakly coupled channels at the threshold of the channels which are the main components of the state~\cite{Wang:2018djr,Wang:2017mrt,Dai:2018tgo,Dai:2018nmw}. The intermediate $P_{cs}(4459)$ only plays an important role above the 4160~MeV, which can be easily understood according to the Dalitz plot of Fig.~\ref{fig:dalitz}, and will shift the bump position of the $X(4160)$ to high energies. In Fig.~\ref{fig:dw_jpsiphi_beta}, we also show the $J/\psi\phi$ invariant mass distribution for the different values of $\alpha$, and one can easily find the bump of $X(4160)$ is clearer for the larger $\alpha$.

\begin{figure}[htbp]
\centering
\includegraphics[width=7cm]{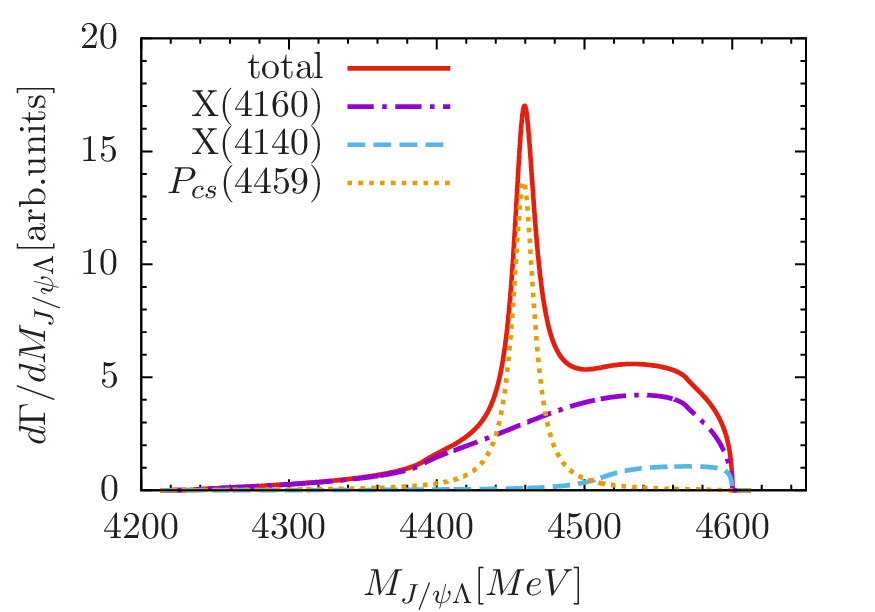}
\caption{The $J/\psi\Lambda$ invariant mass distribution of the $\Lambda_b\to J/\psi\phi\Lambda$ process. The explanations of the curves are the same as those of Fig.~\ref{fig:dw_jpsiphi}.}
\label{fig:dw_jpsiLambda}
\end{figure}

\begin{figure}[htbp]
\centering
\includegraphics[width=6.5cm]{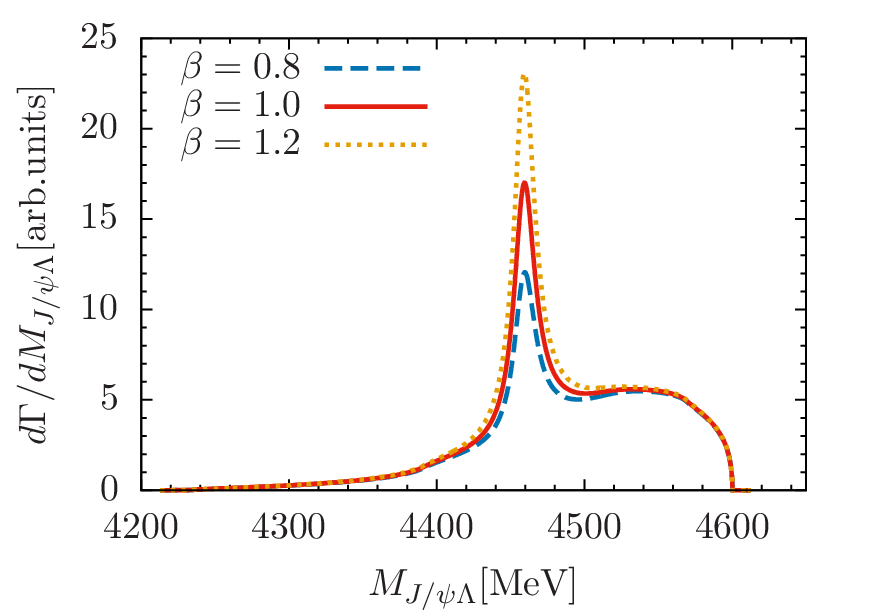}
\caption{The $J/\psi\Lambda$ invariant mass distribution of the $\Lambda_b\to J/\psi\phi\Lambda$ process with different values of $\beta$ ($=0.8,~1.0,~1.2$).}
\label{fig:dw_jpsiLambda_beta}
\end{figure}

In Fig.~\ref{fig:dw_jpsiLambda}, we show the $J/\psi\Lambda$ invariant mass distribution of the $\Lambda_b\to J/\psi \Lambda \phi$, and predict a clear peak around 4460~MeV, associated to the hidden-charm pentaquark $P_{cs}(4459)$. The  $X(4140)$ and $X(4160)$ resonances only have contribution in the region of $M_{J/\psi\Lambda}>4500$~MeV and $M_{J/\psi\Lambda}>4400$~MeV, respectively,  and they do not significantly affect the peak position of the $P_{cs}(4459)$, which can be understood through the Dalitz plot of Fig.~\ref{fig:dalitz}. We also show the $J/\psi\Lambda$ invariant mass distribution with the different values of $\beta$, the relative production weight of the $P_{cs}(4459)$, in Fig.~\ref{fig:dw_jpsiLambda_beta}. It should be pointed out that we do not know the exact value of $\beta$, which should be determined by comparing with the measurements. Since the $P_{cs}(4459)$ was observed in the $J/\psi\Lambda$ invariant mass spectrum of the $\Xi_b\to J/\psi\Lambda K$, it is expected that there is a signal of $P_{cs}(4459)$  in the  $J/\psi\Lambda$ invariant mass distribution of $\Lambda_b\to J/\psi\Lambda\phi$.
 To confirm the existence of the $P_{cs}(4459)$, it is necessary to observe this state in another different process, which would strengthen its interpretation as exotic state~\cite{Chen:2020uif,Peng:2020hql,Wang:2020eep}, rather than possible kinematical effects.

\begin{figure}[htbp]
\centering
\includegraphics[width=6.5cm]{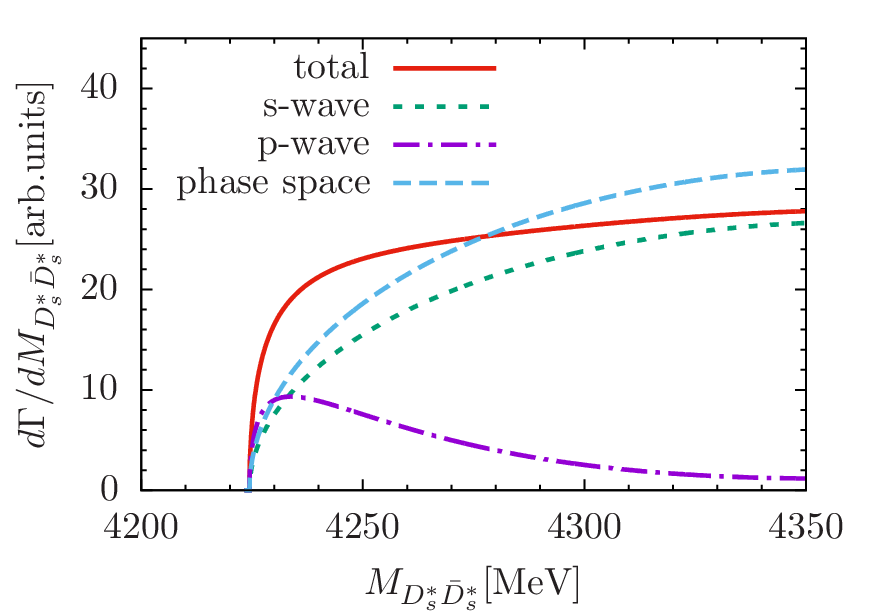}
\caption{The $D_s^*\bar{D}_s^*$ invariant mass distribution of the $\Lambda_b\to D^*_s\bar{D}^*_s\Lambda$ process. The curves labeled as `$S$-wave' and `$P$-wave' show the contributions from the $S$-wave of Eq.~(\ref{eq:swave}) and the $P$-wave of Eq.~(\ref{eq:pwave}). The curve labeled as `total' corresponds to the total contributions of Eq.~(\ref{eq:dw_dsds}), and the curve labeled as `phase space' is the phase space distribution, which is normalized to the same area from threshold to 4350~MeV.}
\label{fig:dw_Dsstar}
\end{figure}

\begin{figure}[htbp]
\centering
\includegraphics[width=7cm]{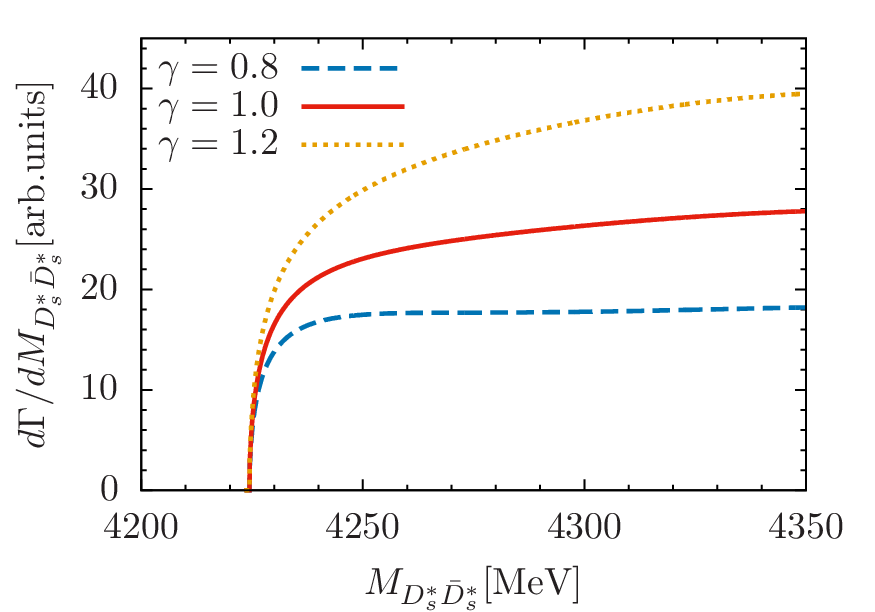}
\caption{The $D_s^*\bar{D}_s^*$ invariant mass distribution of the $\Lambda_b\to D^*_s\bar{D}^*_s\Lambda$ process with different values of $\gamma$.}
\label{fig:7}
\end{figure}

As we mentioned above, the signal of the bump structure from the $X(4160)$ is usually difficult to be identified in the $J/\psi\phi$ mass spectrum, because the width of the $X(4160)$ is too large and its mass is close to the one of the $X(4140)$. Searching for the signal of the $X(4160)$ in other channels is also important. Although the $X(4160)$ mass is less than the $D^*_s\bar{D}^*_s$ threshold, we expect it causes an enhancement structure near the $D_s^*\bar{D}_s^*$ threshold in the  $D_s^*\bar{D}_s^*$ mass spectrum, since the $X(4160)$ is predicted to couple strongly to the $D_s^*\bar{D}_s^*$ channel within the hidden gauge formalism~\cite{Molina:2009ct}. In Fig.~\ref{fig:dw_Dsstar}, we present the $D_s^*\bar{D}_s^*$ invariant mass distribution of the $\Lambda_b\to D^*_s\bar{D}^*_s\Lambda$ process with $\gamma=1$. There is a significant enhancement close to the threshold with respect to the phase space distribution, which is the reflection of the $X(4160)$ and should not be misidentified with a new resonance. We also show the results  for the different values of $\gamma$ in Fig.~\ref{fig:7}, where one can see that the enhancement structure close to the threshold is always clear for the different values of $\gamma$.  Our prediction can be tested by future measurements about the $\Lambda_b\to D{_{s}^{*}}\bar{D}{_{s}^{*}}\Lambda$ reaction.

\section{Conclusions}
\label{sec:conc}
In this work, we have investigated the $\Lambda_b\to J/\psi\Lambda\phi$, by taking into account the $X(4140)$, $X(4160)$, and $P_{cs}(4459)$, and including the mechanisms of the $W^-$ internal and external emissions. In the $J/\psi\phi$ mass distribution, one can find  a clear peak of $X(4140)$, and a bump structure around 4160~MeV followed by a cusp structure around $D^*_s\bar{D}^*_s$ threshold, which can be associated to the $X(4160)$.  We also expect that the hidden-charm pentaquark with strangeness $P_{cs}(4459)$, recently observed by the LHCb Collaboration, causes a clear peak in the $J/\psi\Lambda$ invariant mass distribution.  
It should be noted that it is necessary to confirm this state in another process, which would strengthen its interpretation as exotic state, rather than possible kinematical effects.

Indeed, the signal of the bump structure from the $X(4160)$ is usually difficult to be identified in the $J/\psi\phi$ mass spectrum with the low data sample, because the width of the $X(4160)$ is too large, and its mass is close to the one of the $X(4140)$. Although the $X(4160)$ mass is less than the $D^*_s\bar{D}^*_s$ threshold,
we have predicted that there is an enhancement structure near the $D_s^*\bar{D}_s^*$ threshold in the  $D_s^*\bar{D}_s^*$ mass spectrum, since the $X(4160)$ is predicted to couple strongly to the $D_s^*\bar{D}_s^*$ channel with in the hidden gauge formalism~\cite{Molina:2009ct}.

In summary, we strongly suggest further measurements of these two processes to clarify the nature of $X(4140)$ and $X(4160)$ resonances, and to confirm the existence of the $P_{cs}(4459)$.

\begin{acknowledgments}
This work is partly supported by the National Natural Science Foundation of China under Grants No. 11947089.  It is also supported by the Key Research Projects of Henan Higher Education Institutions under No. 20A140027, the Project of Youth Backbone Teachers of Colleges and Universities of Henan Province (2020GGJS017), the Natural Science Foundation of Henan (212300410123), the Fundamental Research Cultivation Fund for Young Teachers of Zhengzhou University (JC202041042), and the Academic Improvement Project of Zhengzhou University.
\end{acknowledgments}

\end{document}